\documentclass[fleqn,10pt]{wlscirep}
\usepackage[utf8]{inputenc}
\usepackage[T1]{fontenc}
\title{Chondrule formation by collisions of planetesimals containing volatiles triggered by Jupiter's formation}

\author[1,*+]{S. Sirono}
\author[2,+]{D. Turrini}
\affil[1]{Graduate School of Earth and Environmental Sciences, Nagoya University, Nagoya, Japan}
\affil[2]{Turin Astrophysical Observatory, National Institute of Astrophysics (INAF), Pino Torinese, Italy}

\affil[*]{sirono@eps.nagoya-u.ac.jp}

\affil[+]{these authors contributed equally to this work}

\begin{abstract}
Chondrules are spherical or subspherical particles of crystallized or partially crystallized liquid silicates that constitute large-volume fractions of most chondritic meteorites. Chondrules typically range $0.1-2\,$mm in size and solidified with cooling rates of $10-1000\,{\rm K\,h^{-1}}$, yet these characteristics prove difficult to reconcile with proposed formation models. We numerically show that collisions among planetesimals containing volatile material naturally explain both the sizes and cooling rates of chondrules. We show that the high-velocity collisions with volatile-rich planetesimals first induced in the solar nebula by Jupiter's formation produced increasing amounts of silicate melt for increasing impact velocities above $2\,{\rm km\,s^{-1}}$. We propose that the expanding gas formed from volatile materials by collisional heating dispersed and cooled the silicate melt, resulting in droplet sizes and cooling rates consistent with the observed sizes and inferred cooling rates. We further show that the peak melt production is linked to the onset of Jupiter's runaway gas accretion, and argue that the peak age of chondrules points to Jupiter's birth dating 1.8 Myr after CAIs.
\end{abstract}
\begin{document}


\flushbottom
\maketitle
%
%
\thispagestyle{empty}


\section*{Introduction}
Chondrules are spherical particles of crystallized liquid silicates that constitute large-volume fractions of chondritic meteorites. Chondrules typically range 0.1--2\,mm in size\cite{Friedrich} and solidified with estimated cooling rates\cite{cool} of $10-1000\,{\rm K\,h}^{-1}$. The widespread presence, high volume fraction (exceeding 80\% in ordinary chondrites) and spherical (or subspherical) shape of chondrules suggest that they originated from an unknown melting process occurring ubiquitously in the solar nebula. The $^{26}$Al-$^{26}$Mg age of the oldest chondrules dates the time of this ubiquitous melting process to $1.8\,$Myr after CAIs  (Ca--Al-rich inclusions) using the canonical $^{26}$Al/$^{27}$Al value\cite{Pape}. Because of their fundamental role in planet formation, collisions between planetesimals are such a ubiquitous process and the collisional production of melt droplets has been proposed as the chondrule formation mechanism since the early 1950s\cite{Urey, Asphaug, Sanders}. The environmental conditions of the solar nebula required to collisionally form chondrules, however, are still a matter of debate.

Low-velocity collisions like those occurring in unperturbed protoplanetary disks ($\sim 100\,{\rm m\,s}^{-1}$) can produce silicate melt droplets if the interior of the impacting planetesimals is molten\cite{Zook, Asphaug, Sanders}. While the existence of early-formed differentiated planetesimals in the inner Solar System is confirmed by meteoritic data\cite{lichtenberg2023}, melt droplets resulting from their impacts may form large chondrules because the surrounding nebular gas is too rarefied to break them down to millimeter sizes\cite{Friedrich}. If the nebular gas is sufficiently dense, melt droplets of small sizes can be formed. In planetesimal disks characterized by high collisional velocities ($> 2.5\,{\rm km\,s}^{-1}$)\cite{Johnson} the impact jetting process\cite{Johnson} can produce melt droplets consistent with the observed sizes and cooling rates. However, the favoured impact jetting scenario requires the widespread presence of massive planetary embryos\cite{Johnson,hasegawa2016}, whose escape velocities can cause impacts at velocities above $2.5\,{\rm km\,s}^{-1}$, and is more effective in producing chondrules in the orbital region of the terrestrial planets than in the asteroid belt\cite{Johnson,hasegawa2016}. Recent work indicates that impact jetting can produce chondrules also across the outer Solar System\cite{Cashion2025} and can result in the in situ production of carbonaceous chondrules (CC). However, this process requires extended growth times of Jupiter's core since the onset of its runaway gas accretion promptly halts chondrule production by impact jetting\cite{Cashion2025}.


Recent observational and theoretical results reveal that protoplanetary disks cross primordial phases of dynamical and collisional excitation of their planetesimals when their first massive planets form\cite{Turrini19,Testi2022,Bernabo2022}. Collisional studies consistently show that this process is effective in producing high velocity collisions among primordial populations of planetesimals even in the absence of planetary embryos\cite{Turrini12,Turrini19,carter2020}. The comparison of disk population studies with the estimated enhancement in the abundance of disk dust caused by these collisions point to typical formation timescales of 1-2 Myr for giant planets\cite{Bernabo2022}, in contrast with the requirement of extended formation timescales of Jupiter's core to sustain impact jetting in the outer Solar System\cite{Cashion2025}. In the inner Solar System the asteroid belt is the first orbital region to become collisionally excited as a result of Jupiter's formation\cite{Turrini12,carter2020}, with the intensity of its collisional evolution being linked to Jupiter's migration\cite{Turrini12,carter2020}. Jupiter's appearance also triggers the first injection of volatile-rich planetesimals inward of the water snowline\cite{Turrini11,Turrini14,Turrini18}, the magnitude of this process also depending on Jupiter's migration\cite{Turrini14,Raymond17,Pirani19}.

In this work we show that the Jupiter-driven dynamical excitation of volatile-rich planetesimals capable of high-velocity impacts on planetary bodies in the asteroid belt\cite{Turrini11,Turrini14} makes it possible to reproduce the chondrule size range even in case the target planetesimals are not characterized by a fully molten interior due to heating by decay of $^{26}$Al\cite{Lichtenberg18}. Specifically, the sublimation and expansion of volatile elements in the impacts make it possible for the melt droplets to break down to the observed millimeter sizes. To this end we conducted detailed numerical simulations of the planetesimal dynamical excitation induced by Jupiter's formation using the Mercury-Ar$\chi$es parallel n-body code\cite{Turrini19,Turrini21}, and characterized the planetesimal collisional environment using statistical collisional methods\cite{OBrien11,Turrini19} (see Methods). Because of the uncertainty on Jupiter's formation and migration history, we follow the approach of previous studies\cite{Alibert18,carter2020} and consider both the classical scenario of in situ formation\cite{Alibert18,Aguichine22} and a scenario with extensive migration where Jupiter starts its growth beyond the N$_{2}$ snowline\cite{Pirani19,Oberg19} (see Methods).

\section*{Formation of melt by individual planetesimal collisions}

To quantify the amount of silicate melt produced by the planetesimal impacts and inform our global melt production computations, we simulated head-on collisions of two spherical planetesimals using the iSALE shock-physics code\cite{iSALE,Davison}. The reference simulations involve planetesimals of 100 and 400\,km in diameter, respectively, although we tested multiple combinations of target and impactor sizes (see Methods). The impacting planetesimals are described by the equation of state of dunite for simplicity. As the addition of large amounts of volatile materials (especially water) can reduce the amount of melt, in analyzing our results we regard our estimates of the melt production as upper bounds and account for a possible reduction\cite{Cashion21} up to a factor of 10 when the water mass fraction is 20\%. As we discuss in the following, this inhibiting role of volatiles in the collisional production of melt could have played a key role in the petrogenesis of the different classes of chondrites. On the other hand, the amount of melt can be larger than in our simulations if the interior of the target planetesimal is molten as expected in the case of the earliest generations of planetesimals in the inner Solar System\cite{lichtenberg2023}. The full set of input parameters for the impact simulations are shown in the Methods.

The impact velocity is varied between 1 and 10\,km\,s$^{-1}$ based on our n-body simulations (see Methods). The porosity of the planetesimals depends on whether and to which level they experience sintering, melting and compaction, which in turn depends on the formation time and place of the planetesimals\cite{lichtenberg2023}. As our main investigation focuses on target planetesimals formed in the inner solar system and impactors formed in the outer solar system (see Methods), in the iSALE simulations the porosity of the 100\,km impactors is varied between 0.1 and 0.9 under the assumption they underwent moderate to no melting due to the offset in the planetesimal accretion time between inner and outer Solar System\cite{lichtenberg2023}. The target planetesimals with a diameter of 400\,km are assumed to have zero porosity to represent internally-evolved planetary bodies\cite{lichtenberg2023}, consistent with the density data for the differentiated asteroid Vesta provided by the HED (Howardite–Eucrite–Diogenite) meteorites and the NASA mission Dawn\cite{Consolmagno}. The target planetesimal would be volatile-poor due to the likelihood of volatile loss during low-temperature thermal metamorphism\cite{Newcombe, Grewal, Grewal25} and, depending on its degree of internal evolution\cite{Lichtenberg18,lichtenberg2023}, could be limitedly or highly depleted of iron throughout its mantle because of Fe migration or FeS percolation toward the core\cite{Neumann,lichtenberg2023}, the latter process not requiring the global melting of the planetesimal\cite{Neumann,lichtenberg2023}.

Figure~\ref{snap}a shows a time sequence of the evolution of the melt production in a collision with an impact velocity of $5$\,km\,s$^{-1}$ between 100 and 400\, km diameter planetesimals with an impactor porosity of $0.4$. Even at an initial temperature of $200$\,K characteristic of equilibrium with the nebula gas at about $2$\,AU (see Methods), the surface around the impact point is melted, and the melt fraction approaches unity. Because the ANEOS equation of state does not include the phase transition (i.e., melting and vaporizing), the melt fraction is estimated from entropy\cite{Kurosawa}, The entropy at the incipient and complete melting at $10^5\,$Pa are $2.35\,{\rm kJ\,K^{-1}\,kg^{-1}}$ at $1373\,$K and $3.31\,{\rm kJ\,K^{-1}\,kg^{-1}}$ at $2173$\,K, respectively. Thus, the melt fraction of $0.5$ roughly corresponds to $(1373+2173)/2\simeq 1800\,$K. The melt is produced around the impact site and spreads laterally as the target deforms. When $t=75\,$s, the amount of the melt reaches a maximum of $80$\,\% of the impactor mass. The thickness of the melted layer around the impact point is $\simeq 35$\,km at $25$\,s and decreases to $10$\,km at $75$\,s after the collision. The melt layer expands horizontally as the target planetesimal breaks up. Figure~\ref{snap}b shows the total amount of melt produced in a collision as a function of the impact velocity. If the impact velocity is higher than $6\,{\rm km\,s}^{-1}$, the melt mass exceeds that of the impactor, reaching up to a few percent of the mass of the target. The impactor porosity dependence is shown in Fig.~\ref{snap}c. It can be seen that the amount of melt increases as porosity increases, and is twice that of $0.4$ at $0.7$. The porosity of $0.4$ is adopted based on the density data for the Saturnian satellite Phoebe from the NASA mission Cassini\cite{Porco2005} but could be lower for impactors that are thermally evolved. The normalized amount of melt decreases to $0.12$ for zero porosity from $0.53$ for porosity of $0.4$ at an impact velocity of $4\,{\rm km\,s^{-1}}$.


\section*{Global melt production caused by Jupiter's formation}

By combining the melt production by individual impacts as a function of the impact velocity simulated with iSALE (Fig.~\ref{snap}b) with the average impact velocities and the number of impacts computed by processing the $n$-body simulations (see Methods) we characterize the average melt production rate across the inner Solar System as a function of time in both Jupiter's in situ formation and extensive migration scenarios (Fig.~\ref{Fig1}). As presented above, the melt production we adopt is the result of the vertical collision obtained by iSALE simulation, where volatile material is not included. Therefore, the global amount of the melt determined from processing the n-body simulations should be regarded as an upper bound.

The results show that the timing of peak melt production immediately follows the onset of the runaway gas accretion onto Jupiter, consistently with previous studies showing that peak collisional excitation is reached about $0.1$~Myr after runaway gas accretion begins\cite{Turrini12,Turrini18}. Melt production by impactors originating beyond the water snowline is efficient between 2 and 4 AU with the region between 3 and 4 AU being characterized by more intense and longer production with respect to the region between 2 and 3 AU (Figure~~\ref{Fig1}). The scenario with extensive migration produces about $0.2$~M$_\oplus$ of melt (see inset in Figure~~\ref{Fig1}, right plot) and proves about an order of magnitude more efficient than the in situ formation scenario, which results in a total melt production of about $0.01$~M$_\oplus$ (see inset in Figure~~\ref{Fig1}, left plot). The orbital region where collisional melt production is effective is consistent with the indications of origins close or beyond the water snowline from the oxidation states of the parent bodies of both carbonaceous and non-carbonaceous planetesimals\cite{Grewal2024} as we will discuss when presenting the possible petrogenesis of the different classes of chondrites.

Both Jovian formation scenarios produce masses of melt large enough to reproduce the present-day asteroid belt after accounting for the mass loss it experienced since its formation\cite{weidenschilling2011} (see Figure~~\ref{Fig1} and Methods) and the uncertainties in the collisional melt production efficiency we discussed above. While not the direct focus of this study, our simulations also show that both Jovian formation scenarios result in compositional structures of the asteroid belt globally compatible with the present one, albeit they predict different origins for the volatile-rich asteroids (see Supplementary Information and Fig. S1). 


In the in situ formation scenario about 1\% of the melt mass is produced by the dynamical excitation caused by Jupiter's core growth before the beginning of the runaway gas accretion (see inset in Figure~~\ref{Fig1}, left plot). Specifically, melt production begins when Jupiter's core is sufficiently massive (10--15\,M$_\oplus$) to excite the nearest planetesimals\cite{Alibert18} and extends for about 0.5 Myr after the peak associated with the runaway gas accretion. Such core-driven production would be consistent with the existence of chondrules older than the peak at about 1.8 Myr suggested by the Pb$-$Pb dating\cite{Pape}.
In the extensive migration scenario melt production by volatile-rich planetesimals has a sharp start when Jupiter nears its current orbits at the end of its growth and migration (see inset in Figure~~\ref{Fig1}, right plot) and also lasts for about 0.5 Myr after the initial peak. In this scenario, the existence of older chondrules could be explained by the impacts between the first generation of differentiated planetesimals in the inner Solar System\cite{lichtenberg2023}. Their origins in impacts not involving water-rich planetesimals could explain the unusual large sizes of these possibly older chondrules\cite{Pape}.

While our results show that substantial amounts of chondrules are formed between $1.8$ and $2.3$\,Myr, this temporal window is shorter than that spanned by chondrule ages\cite{Pape,lichtenberg2023}. The formation of Saturn\cite{Coradini2011,Turrini12,Ronnet2018} can extend the duration of the collisional melt production by volatile-rich planetesimals in the asteroid belt, as it excites the planetesimals in the outer Solar System after the effects triggered by Jupiter's formation end. The implantation of carbonaceous planetesimals in the asteroid belt by Jupiter (see Supplementary Information) naturally creates the conditions for the extended formation times of the carbonaceous chondrites\cite{Pape,lichtenberg2023} under the effects of Saturn's formation. Further mechanisms that can sustain the dynamical excitation of the inner Solar System beyond what is caused by Jupiter's formation are discussed in the Supplementary Information. 

\section*{Expansion of melt layer by volatiles}
The planetesimals acting as high-velocity impactors originate beyond the water snow line (between 3 and 7 AU in the in situ formation scenario and between 3 and 30 AU in the extensive migration scenario) and, depending on their specific formation region, contain volatile materials in the form of hydrated minerals\cite{Robert}, ices\cite{Turrini14,Raymond17} and organic materials\cite{Bergin2015,Turrini21}. The expected abundance of H$_2$O, the most abundant of such volatile materials\cite{Oberg19}, ranges from 10 wt.$\%$ to 30 wt.$\%$\cite{Robert,Turrini21}. When a collision involving one of these planetesimals produces silicate melt, the volatile materials in the melt can expand quickly and cause the melt to accelerate and to form droplets. The role of outgassing volatiles in accelerating solid particles like dust grains is observationally supported by the Deep Impact experiment on Comet 9P/Tempel 1\cite{Holsapple, Jorda, Keller}.


The volatile material becomes gaseous and quickly expands upward. This process is modeled as a 1-D shock tube problem, as shown schematically in Figs.~\ref{Sche}a and b. Initially, there is a layer of silicate melt at rest with a temperature $T_0$ and thickness $L_0$ containing volatile material of molecular weight $m$ ($m=18m_{\rm H}$ for H$_2$O, where $m_{\rm H}$ is the hydrogen mass). 

The mass fraction of the volatile material is $f$. The volatile material expands into the ambient nebular gas at a temperature $T_{\infty}=200\,$K and gas density $\rho_{\infty}=2\times 10^{-7}\,{\rm kg\,m}^{-3}$, which correspond to the values at about 2\,AU in the adopted solar nebula model (see Methods). The gas is treated as an ideal gas for simplicity. The thickness $L_0$ depends on the impact velocity and impactor size. If the target planetesimal is melted, $L_0$ is determined by the spatial extent of the mixing of the impactor and the molten interior. We adopted spherical symmetry because the gas and melt expand radially when the cooling proceeds, at much later times than Fig.1~a.

If the dynamic pressure of the flowing gas is higher than the pressure inside a droplet resulting from surface tension, the droplet breaks up. The critical breakup size $D_{\rm c}$ of the droplets is given as
\begin{equation}
  D_{\rm c}={{\rm We_{\rm c}}\sigma \over \rho_{\rm gas}(v_{\rm gas}-v_{\rm melt})^2},
  \label{size}
\end{equation}
where We$_{\rm c}=10$ is the critical Weber number for breakup\cite{Kadono}, $\rho_{\rm gas}$ is the gas density changing with time, and $\sigma=0.065\,{\rm J\,m}^{-2}$ is the surface tension of the melt containing water\cite{Gardner}. The size of a droplet after the breakup was assumed to be half of the critical size\cite{Kadono} given above. Coalescence of the droplets was excluded because large gas pressure\cite{Krishnan} and large viscosity of droplets\cite{Tang} prevent the coalescence.

Typical evolution of the spatial density of melt and gas is shown in Fig.~\ref{Fig3}a. The parameters for this simulation are $m/m_{\rm H}=18$ (H$_2$O), $L_0=10\,$km (Fig.~\ref{snap}a), and $f=0.1$\cite{Rivkin} Both components similarly expand from the surface of the layer. When the melt density decreases to the breakup density\cite{Sparks} (dashed line in Fig.~\ref{Fig3}a), melt droplets are formed. The evolution of the expanding velocity is shown in Fig.~\ref{Fig3}b. The melt and gas layers expand at around the sound speed. The two velocities are almost the same in this figure. The difference in the two velocities is shown in Fig.~\ref{Fig3}c. The velocity difference is $\sim 10\,{\rm m\,s}^{-1}$ at the expanding layer. This velocity difference determines the size of the melt droplet. Figure~\ref{Fig3}d displays the evolution of the melt droplet diameter distribution. The diameter is determined by the  velocity difference between the melt and gas components. The diameter decreases as $r/R_{\rm pla}$ ($R_{\rm pla}$ is the planetesimal radius) increases, reflecting the distribution of the velocity difference shown in Fig.~\ref{Fig3}c. Figure~\ref{sizeevo}a shows the evolution of the average droplet size. The breakup of the melt layer finished at $t/t_0=75$ when the melt volume fraction fell below\cite{Sparks} $0.2$ (Fig.~4a). The average size was constant at approximately $1.1\,$mm.

The expansion of the melt layer proceeds along the horizontal expansion (Fig.~\ref{snap}a). The expansion timescale for the breakup is $83\,$s (in normalized unit $t/t_0=75$), and the horizontal expansion timescale to the maximum melt mass is $75$\,s. These timescales are comparable. The formation of droplets proceeds along the horizontal expansion. The thickness of the melt layer $L_0$ is maximum at the center and decreases as the distance from the center increases, and the thickness shrinks with time. The size distribution of droplets is the superposition of droplets formed at various positions.

Figure~\ref{sizeevo}b shows the evolution of the temperature decrease $\Delta T$ of the melt droplets averaged over $x$ for the same parameters used in Fig.~4. The temperature decrease exceeds $400\,$K (assumed solidus temperature) at $t_{\rm c}=0.68\,$h. The dashed line in Fig.~\ref{sizeevo}b is the semi-analytical solution given by Eq.~(\ref{DT}), which nicely explains $\Delta T(t)$. The inset of Fig.~\ref{sizeevo}b shows the evolution of the cooling rate $|dT/dt|(t)$. The cooling rate is fast at the beginning and slows down. At $t=0.68\,$h, $|dT/dt|(t)=160\,{\rm K\,h}^{-1}$. The cooling rate is also well explained by Eq.~(\ref{dTdt3}). On the other hand, the average cooling rate is given by $\Delta T(t_{\rm c})/t_{\rm c}=660\,{\rm K\,h}^{-1}$ at $t=0.68\,$h. At this time, the gas and melt spatial densities are $1.0\times 10^{-3}\,{\rm kg\,m}^{-3}$ and $3.1\times 10^{-2}\,{\rm kg\,m}^{-3}$ (Fig.~\ref{sizeevo}c, d), respectively, and the gas pressure is $7.3\times 10^{-3}\,$bar, enough to prevent the isotopic fractionation of chondrules\cite{Cuzzi}. Because the opacity due to melt droplets is still large, radiative cooling\cite{Dullemond} does not proceed until $\Delta T=400\,$K.

We have conducted simulations by varying the molecular weight $m/m_{\rm H}$, the gas mass fraction $f$, and  layer thickness $L_0$ ($T_0=1800$\,K is fixed). Figure~6 shows the parameter dependence obtained from numerical simulations. We determined the average cooling rate at $\Delta T=400\,$K (Fig.~3c). The droplet diameters $D$ for different parameters are compared. The simulation results for droplet size $D$ and the cooling rate $\Delta T/t_{\rm c}$ are well expressed by the semi-analytical solutions [see Eqs.~(\ref{Dfit2}) and (\ref{dTdt3}) in Methods]. The ranges of the size and cooling rate are between $0.38-1.8$\,mm and $28-18,000\,{\rm K\,h}^{-1}$, respectively, which broadly overlap the observed ranges of $0.1-2$\,mm and $1-10^3\,{\rm K\,h}^{-1}$, respectively. The volatile mass fraction $f$ and molecular weight $m/m_{\rm H}$ are constrained as $f<0.1$ and $m/m_{\rm H}>18$, respectively, based on the observed cooling rate of $1-10^3\,{\rm K\,h}^{-1}$. The dependence of the cooling rate on $L_0$, which depends on the planetesimal size and impact velocity, is weak. The chondrule diameter $D$ can be explained within the range of the parameters used in this simulation. In this simulation, the size distribution of melt droplets after breakup was neglected for simplicity. Experimental results\cite{Kadono} showed that droplets much smaller than $D_{\rm c}/2$ can be formed. The chondrule diameter shown in Fig.~5 can therefore be regarded as the maximum size of the size distribution.

\section*{Petrogenesis of non-carbonacous chondrites}

The formation of chondrules by impacts involving volatile-rich planetesimals is in agreement with the enhancement of water advocated by multiple authors to explain the chondrule properties. The variation of the oxygen isotope ($\Delta^{17}$O) in chondrules can be explained by the variation in water content\cite{Williams}. The redox condition and low level of isotope fractionation is consistent with water enhancements by a factor of $\sim 500$\cite{Fedkin12}. The coexistence of type I (FeO-poor) and type II (FeO-rich) chondrules in the same meteorite could be explained by spatial heterogeneity of water. When an impactor hits the target planetesimal, the water in the expanding gas mixes with the target material, while the mixing of target and impactor materials may be inhibited by the interplay between the melt suppression in the impactor due to its volatile content and the different coupling to the gas of the melt droplets from the target and the solid ejecta from the impactor. The mixing with water is not perfect and the amount of surviving water depends on the impact velocity\cite{Turrini14,Turrini18}, so the resulting water-rich material forms type II chondrule and water-poor material forms type I chondrule\cite{Fedkin12}.

Recent work\cite{Grewal} points to the earliest formed non-carbonaceous (NC) planetesimals being oxidized and likely containing water to some extent. Based on these results, we argue that the NC planetesimals whose impact melt formed H-L-LL chondrules plausibly sampled the orbital region inward to the water snowline where nebular temperatures below $400$~K allowed for the presence of phyllosilicates and hydrated rocks\cite{Woitke2018}. Depending on their formation times, the thermal histories of the NC planetesimals could have resulted in their devolatilization or FeS percolation\cite{lichtenberg2023} if not global melting. Devolatilization would have reduce the amount of O available to form FeO from Fe in impact melt while FeS percolation would remove Fe from the outer layers of the NC planetesimals and enrich their inner layers without implying global melting\cite{lichtenberg2023}. Impacts on planetesimals with different abundances of Fe in the melt-producing layers would produce NC chondrules with different amounts of Fe. The combination of these two factors results in the formation of NC chondrules with different amounts of Fe and varying proportions of FeO and suggests that the parent bodies of L-LL chondrules formed earlier than those of H chondrules and underwent higher degrees of oxidation and Fe migration due to greater thermal processing.

Although we did not explicitly model their role, organic materials are the next major component of volatile materials\cite{Turrini21}, possibly being incorporated into planetesimals already inside the water snowline\cite{Bergin2015}. The presence of dry organic materials can result in the formation of reducing gas during collisions and is one of the proposed routes to form the FeO-poor enstatite chondrites\cite{Connolly,Jacquet2018}. Dry organic materials are argued to be present in nebular regions with temperature comprised between 400 and 250 K \cite{Gail2017}, which is compatible with the ambient temperature of the formation region of enstatite chondrites inferred from their abundance of Cl\cite{Marrocchi2024}. Such disk regions would be located inward of the water snow line and, in our disk model, would encompass the asteroid belt. While we do not model their mutual impacts directly, planetesimals excited by the appearance of the 2:1 mean motion resonances with Jupiter represent the main source of high-velocity impactors in the inner asteroid belt\cite{Turrini12,Turrini18}. Impacts in the inner asteroid belt with planetesimals excited by the 2:1 resonance with Jupiter can therefore provide a viable path for the formation of enstatite chondrites in the collisional scenario we explore here. The age of chondrules in an enstatite chondrite\cite{Zhu} of $1.6$\,Myr after CAI might correspond to the first peak of the red curve ($1-2$\,AU in Fig.~2) at $1.7$\,Myr.

Overall, our results suggest that the E-H-L-LL chondrules originate from the early generation of NC planetesimals sampling the progressively colder orbital regions where anhydrous rocks characteristic of the terrestrial planet region would be gradually enriched  in dry organics (400-250 K)\cite{Gail2017}, phyllosilicates (400-200 K) and water (150 K)\cite{Woitke2018}.

\section*{Volatile-rich planetesimals and the NC-CC dichotomy}
The scenario of chondrule production by impacts of volatile-rich planetesimals dynamically excited by Jupiter's formation is consistent with the most recent results on the dichotomy between carbonaceous and non-carbonaceous isotopic families of planetesimals in the asteroid belt at the time of the peak chondrule formation\cite{Kruijer}. 
Recent work suggests that CC chondrules originate from a mixture of NC and CI material\cite{Bryson2021, Onyett2024, Marrocchi2024} and argue in favor of spatially and temporally localized processes rather than disk-wide processes or transport\cite{Bryson2021, Onyett2024}. The mixing of NC and CI material is naturally explained in the collisional scenario we present by the fact that the formation of melt in the volatile-rich impacting body would be suppressed by the presence of the volatiles by up to an order of magnitude with respect to the rock-dominated target body\cite{Cashion21}, making the melt from the target the dominant component of the total melt production. The impacting body would still eject dust and collisional debris, which provide the CI material needed to produce CC chondrules from NC material. As dust ejection and debris production are significantly more efficient than melt production (e.g.by comparing the mass of melt in this work with the collisional dust production in protoplanetary disks\cite{Turrini19, Bernabo2022}), this process is consistent with the estimates of 75-90 wt\% contribution of CI-like material to the formation of CC chondrules from early formed NC chondrules reported by the works cited above. The resulting CI-NC mixture can either condense into planetesimals by streaming instability \cite{Johansen2014} or accumulate on the surface of existing planetesimals by pebble accretion \cite{Johansen2017}: subsequent impacts would then thermally reprocess the mixture into CC chondrules. Furthermore, the production of the bulk of the chondrule population across the asteroid belt following Jupiter’s formation satisfies the requirements of spatially and temporally localized processes proposed\cite{Onyett2024}. In addition, geometric effects not modeled in our study could also have played a role. The study for impact jetting highlights\cite{Wakita2021} how the composition of the jetted material depends on the impact angle. Impact angles below 45° cause the impactor material to be dominant, while at higher impact angles the target material dominates. Similar geometric effects would countribute to explain the absence of mixing between target and impactor droplets in NC chondrules.

The detailed analysis of CB chondrules revealed that collisions with high water content ($\sim 20$\,\%) and a high cooling rate match the zoning profiles\cite{Fedkin}. This is consistent with our results of $f$ dependence of the cooling rate (Fig.~\ref{matome}d) and the implantation of carbonaceous planetesimals from the outer Solar System in the asteroid belt by the interplay between Jupiter's formation and gas drag. To further explore this, we ran additional collisional melt production simulations (see Methods) to assess how accounting for the mutual impacts among planetesimals formed in the outer Solar System affects our results in the orbital region between 1 and 5 AU. In Jupiter's in situ formation scenario, the effects of such impacts over the timespan of our simulations are limited and increase melt production only by a few per cent. In Jupiter's extended migration scenario these impacts can double the total melt production to about $0.4$~M$_\oplus$. However, this estimate does not account for the melt inhibiting effects of volatiles\cite{Cashion21}, which is particularly important for these impacts as it affects both target and impactor. Decreasing the efficiency of melt production by an order of magnitude for these impacts returns a total melt production of $0.22$ ~M$_\oplus$, i.e. an increase of about 10\%. The extended duration of the collisional excitation of the asteroid belt that is caused by Saturn's formation or by the appearance of massive planetary embryos (see Supplementary Information for discussion) would extend the temporal window where these impacts contribute to the production of chondrules and offer a viable path to form CB chondrules $\sim3.8$\,Myr after CAIs\cite{Wolfer2023}. Furthermore, the destruction of implanted volatile-rich planetesimals and the incorporation of the resulting material into later-formed chondritic planetesimals is consistent with the discovery of a cometary xenolith in the matrix of a CR chondrite\cite{Nittler2019}.

\section*{Implications for the formation of Jupiter and the evolution of the Solar Nebula}
The agreement between the observed physical characteristics and abundance of chondrules and those resulting from our melt production model supports the causal connection between chondrule production and the formation of Jupiter. The sharp and marked growth of the Jovian-induced melt production \text{following the onset of Jupiter's runaway gas accretion} (see Fig. \ref{Fig1}) allows to use the measured age of peak chondrule production at 1.8 Myr to accurately date Jupiter's birth for the first time. The appearance of Jupiter at 1.8 Myr in the Solar System agrees with the formation timescale of 1-2 Myr of giant planets in protoplanetary disks estimated from their rise in dust content\cite{Testi2022,Bernabo2022} and suggests that planet formation in our Solar System occurred under environmental conditions common enough in our galaxy. This, in turn, indicates that the study of chondrules and chondritic meteorites in the Solar System provides us insights on the collisional environments existing in protoplanetary disks between the appearance of massive planets and the dissipation of the nebular gas.

The produced chondrules should float in the solar nebula, and eventually, they should either be re-accreted by their parent body or coalesce into a new condritic planetary body\cite{Lichtenberg18,Johansen2017}. The floating time corresponds to the time difference between the formation of the chondrules and either of these events. This time difference is $0-0.3\,$Myr based on the thermal history of the parent bodies of H, L, chondrites, and acapulco-Lodran primitive achondrite\cite{Henke, Gail, Neumann18, Trieloff}. The existence of compound chondrule\cite{compound} indicates that at least a few percent of chondrules floating in the solar nebula encountered splash of chondrules formed by the other planetesimal collisions during the floating time. The floating time should be sufficiently short, otherwise components of different chondrite types would efficiently mixed together. Based on the scenario we modelled, the matrix component in chondritic meteorites comes from the thermally unprocessed fragments (blue part in Fig.1a) of the impacting planetesimals, the chondrules and matrix mixing together during their floating in the solar nebula. The high volume fraction of matrix in carbonaceous meteorites might come from the large water content which reduces the efficiency of the melt production\cite{Cashion21}.

While we focused our collisional computations on the inner Solar System, Jupiter's formation dynamically and collisionally excites also the orbital region beyond the giant planet (see Figure S2 and\,\cite{Turrini19,Bernabo2022,Polychroni2025} for analogous computations for planet-hosting protoplanetary disks). Collisions among planetesimals beyond the current orbit of Jupiter can in principle produce carbonaceous chondrules in parallel to the formation of non-carbonaceous chondrules by the impacts on the planetesimals originating in the asteroid belt. Once Saturn form, the combined perturbations of the two giant planets would inject and implant the resulting carbonaceous chondrite parent bodies in the asteroid belt\cite{Ronnet2018,Pirani19}. As discussed above, however, because of the ice-rich nature of the planetesimals in the outer Solar System the production of CC melt/chondrules would be damped in favor of that of dust and collisional debris that would be prevented from drifting toward the inner Solar System by the barrier effect of Jupiter and Saturn. Moreover, the lower intrinsic impact probabilities in the outer Solar System resulting from the longer orbital periods/lower spatial density\cite{Turrini19} would further hinder the production of impact melt and chondrules.

Finally, in this work we focused on the production of chondrules by collisions involving volatile-rich planetesimals, yet the physical scenario we describe allows for an additional process to occur. The expansion of the volatile-rich impact plumes generates shock wave that propagate through the surrounding nebular gas and, if the dust-to-gas ratio is sufficiently high, can melt the dust into chondrules \cite{Stewart}. While the chondrule production efficiency of impact plumes is not yet quantified, this process is complementary to the one we investigated, as planetesimal collsions are invoked as the source of the required high dust-to-gas ratio\cite{Stewart}. The interplay between melt-producing collisions and shock-generating impact plumes therefore offers a way to enhance the efficiency of chondrule production.



\section*{Methods}

\subsection*{Dynamical excitation of the planetesimals}

We performed numerical simulations of the dynamical excitation of the planetesimals in the solar nebula induced by Jupiter's formation using the Mercury-Ar$\chi$es parallel $n$-body code\cite{Turrini19,Turrini21}. The planetesimal disk is simulated by test particles distributed uniformly in  semimajor axis with a spatial density of 5000 test particles/AU and evolving under the influence of the forming Jupiter and the nebular gas. The test particles are initially distributed between 1 and 10 AU in Jupiter's in situ formation simulation and between 1 and 36 AU in Jupiter's extensive migration simulation. In both simulations we prevented particles to populate the feeding zones of Jupiter's core to account for its growth process\cite{Turrini19,Turrini21}. Each test particle represents a swarm of planetesimals whose mass is set by the local nebular conditions at its formation region (see below). The initial eccentricities and inclinations (in radians) of the planetesimals are uniformly extracted between 0 and 0.01\cite{Turrini19}. The test particles do not possess gravitational mass, so they do not perturb each other or Jupiter, but they have inertial masses to allow for quantifying the effects of the disk gas on their dynamical evolution (see below). The dynamical evolution of the planetesimal disk is simulated for 3 Myr with timestep of 15 days.

The gas surface density profile of the solar nebula is derived from recent solar nebula models\cite{Alibert} calibrated on the observations of protoplanetary disks and is defined as 
$\Sigma(r) = \Sigma_{0}\left(\frac{r}{50\,{\rm AU}}\right)^{-0.8} \exp\left[-\left(\frac{r}{50\,{\rm AU}}\right)^{1.2}\right]$ 
where $\Sigma_{0} = 36$ g/cm$^{2}$ is the density at the characteristic radius of 50 AU, whose value is set based on the constraints from the current architecture of the solar system\cite{Kretke}. The solar nebula is assumed to be in a steady state and its mass does not change during the n-body simulations. The extension of the planetesimal disk is assumed to coincide with the characteristic radius\cite{Kretke}, while the gas disk is assumed to be a factor of four more extended based on observational constraints\cite{Turrini21}. The total gas mass of the solar nebula amounts to 0.053 M$_\odot$ (0.033 M$_\odot$ within 50 AU). The temperature profile in the midplane of the solar nebula is 
$T(r) = T_{0} \left(r/1\,{\rm AU}\right)^{-\beta}\,{\rm K}$
where $T_{0} = 280$ K\cite{Alibert} and $\beta=0.65$\cite{Oberg19}, resulting in the water ice snowline falling at approximately 3\,AU. 

The disk temperature profile allows to quantify the fraction of heavy elements condensed as rocks and ices at the different radial distances\cite{Turrini21}, i.e. the local dust-to-gas ratio. This value is then increased by a factor 1.6 to account for the spatial concentration of solids with respect to the gas\cite{Turrini21} discussed above. To compute the swarm mass associated with each test particle, we integrate the gas disk profile over a 0.1 AU wide ring centered on the initial orbital position of the relevant particle and multiply the resulting mass by the local dust-to-gas ratio and the concentration factor of solids discussed above, then divide it by the number of test particles within the ring\cite{Turrini21}.

To accurately model the interactions between particles and the nebular gas, Mercury-Ar$\chi$es\cite{Turrini21} includes the treatment of gas drag based on the Reynolds and Mach number of the planetesimals, 
of disk gravity in the axisymmetric disk approximation, 
and of the formation of a gap around the growing giant planet.
To quantify the effects of the nebular gas the test particles are attributed inertial masses by assuming a characteristic diameter of $100$\,km\cite{Johansen2017} 
and by setting their density to 2.4 g\,cm$^{-3}$ if they originate within the water snowline 
and  1.0 g\,cm$^{-3}$ if they originate outside of it\cite{Turrini19}. Since the damping efficiency of gas drag decreases for larger planetesimals, by focusing on 100 km-wide bodies we are conservatively minimizing the amount of dynamical excitation, hence the impact velocities, of the primordial planetesimals in our simulations.

Jupiter's mass growth is modeled using a parametric approach\cite{Turrini21} reproducing the behaviour of realistic growth curves\cite{DAngelo21}. 
Jupiter starts as a Mars-sized embryo of $M_0 =0.1$\,M$_\oplus$ and grows to $M_c = 30$\,$M_\oplus$ (i.e., a critical-mass core of 15\,M$_\oplus$ plus an extended atmosphere of 15\,M$_\oplus$) as $M_{J}=M_{0}+\left( \frac{e}{e-1}\right)\left(M_{c}-M_{0}\right)\left( 1-e^{-t/\tau_{c}} \right)$ over a time $\tau_c = 1.7$ Myr.
As shown by Fig. 1, the specific choice of $\tau_c$ has negligible impact on the total melt production: this specific value is chosen to achieve the best fit to the constraints from the age of chondrules as the simulations show a delay of about 0.1 Myr between the onset of Jupiter's runaway gas accretion and the peak dynamical excitation of the planetesimals. After reaching the critical mass $M_c$, Jupiter begins its runaway gas accretion phase and grows as $ M_{J}=M_{c}+\left( M_{F} - M_{c}\right)\left( 1-e^{-(t-\tau_{c})/\tau_{g}}\right)$ where $M_F=317.9$\,M$_\oplus$ is its final mass (based on the IAU 2009 values of Jupiter's and Earth's masses), and $\tau_{g}$=$10^{5}$ years its runaway gas accretion timescale\cite{DAngelo21}. Following recent theoretical results\cite{Tanaka2020}, the runaway gas accretion process is assumed to stop due to the inability of the solar nebula to supply gas to the gap surrounding the growing Jupiter before the dissipation of the solar nebula itself, meaning that chondrule production and the formation of chondritic planetesimals are possible after the completion of Jupiter's growth.

Jupiter's migration, when included, is modeled after the migration tracks from\,\cite{Mordasini2015} following the parametric approach by\,\cite{Turrini21}. During the core growth phase Jupiter undergoes linear Type I migration regime with drift rate 
$\Delta v_{1} = \frac{1}{2}\frac{\Delta a_{1}}{a_{p}}\frac{\Delta t}{\tau_{p}}v_{p}$
where $\Delta t$ is the timestep of the N-body simulation, $\Delta a_{1}$ is the radial displacement during the first growth phase, and $v_{p}$ and $a_{p}$ are the instantaneous planetary orbital velocity and semi-major axis, respectively. During the runaway gas growth phase, encompassing the transition to full Type I regime first and Type II regime later, the drift rate becomes  
$\Delta v_{2} = \frac{1}{2}\frac{\Delta a_{2}}{a_{p}}\frac{\Delta t}{\tau_{g}}\exp^{-\left(t-\tau_{c}\right)/\tau_{g}} v_{p}$
where $\Delta a_{2}$ is the radial displacement during this second phase. The values of $\Delta a_{1}$ and $\Delta a_{2}$ are set so that Jupiter's core migration accounts for 40\% of its total orbital displacement\cite{Turrini21}.

\subsection*{Collisional evolution of the planetesimals}

Snapshots of the orbital elements of the test particles, including the information on their initial semimajor axes/formation regions, are recorded every $10^{4}$ years. We used well-tested statistical collisional methods developed for the study of the asteroid belt (see\,\cite{OBrien11} and references therein) to characterize the impact probabilities and velocities among the planetesimals at each snapshot. In our reference case we consider as targets those particles with formation regions within 5 AU, i.e. Jupiter's current orbit, while the impactors are those particles that have formation regions beyond the water snowline at 3 AU\cite{Onyett2024} and pericenters below 5 AU. Impact probabilities are computed only among pairs of particles satisfying these conditions to avoid overestimating the number of impact events. For each target body we compute the average impact velocity with all possible impactors and conservatively use this value to characterize the melt production, to avoid being skewed by rare high-speed impact events. Average impact velocities range between 0.5 and 5.5 km/s at peak excitation in Jupiter's in situ formation scenario while they range between 2 and 9 km/s in Jupiter's extensive migration scenario.

We process the impact probabilities and average impact velocities with the Debris code\cite{Turrini19,Bernabo2022} to characterize the collisional evolution of the inner Solar System. Specifically, we convert the impact probabilities into numbers of collisions and associated mass losses following the methodology from\,\cite{Bernabo2022}. We adopt the size-frequency distribution in mass\cite{Turrini19} of planetesimals formed by pebble concentration, characterized by exponential slope $\gamma=1.6$\cite{Krivov}, to resolve each swarm mass across the range of diameters from 100 to 1000 km, i.e. the size range suggested for the primordial asteroid belt\cite{Morbidelli}. The conversion from diameters to masses is performed using the density values discussed for the n-body simulations. We use the masses of the individual planetesimals to quantify the number of planetesimals in each size bin. When resolving the number of impacts between two colliding swarms, target planetesimals with diameter $d_i$ are allowed to collide only with impacting planetesimals of equal or smaller size\cite{Turrini19}. As discussed above, the particles in the n-body simulations have a characteristic {\rm diameter} of 100 km, i.e. the smaller size in the mass spectrum considered in the collisional analysis. As the damping efficiency of gas drag increases at smaller sizes, this means that we are considering the least excited collisional environment and the lowest impact velocities in our melt production computations. We combine the impact characterization by Debris with the results of the impact simulations with iSALE (see below) to quantify the melt production from the collisional evolution of the target swarms across the different snapshots.

To account for the mass evolution of the swarm masses due to impacts\cite{Bernabo2022}, we use the scaling law for collisional mass loss from \cite{Genda} that is valid both in the regime of cratering erosion and that of catastrophic disruption. The cumulative mass loss experienced by the target bodies over all the impacts between two colliding swarms is subtracted from the target swarm mass, as the lost mass is assumed to be converted into collisional debris and large pebbles\cite{Wada2021} that, being more strongly affected by gas drag, are assumed to be efficiently removed from the swarm. The mass of the impactor swarm is instead reduced by the product between the number of impacts and the mass of the impacting bodies. As an example, if a target body of 800 km impacts with three impactors of 100 km, the mass of the target swarm is reduced by the mass lost by the 800 km-wide body while the mass of the impactor swarm is reduced by the mass of the three impactors. The collisional mass loss proves an order of magnitude more efficient than melt production in removing mass from the primordial planetesimal disk, and plays a key role in regulating the intensity and the duration of the melt production. Specifically, larger initial numbers of impacting bodies, as those characterizing the Jovian extensive migration scenario, cause more intense collisional evolution and higher mass loss, which in turn results in a faster decrease in the number of impacts\cite{Bernabo2022}.

To assess the robustness of the results with respect to the choice of target bodies, we performed a second collisional evolution and melt production simulation where we considered as target bodies all planetesimals orbiting inside 5 au, independently on their formation region. This translates into extending the definition of target bodies to include also the volatile-rich planetesimals formed beyond Jupiter's current orbit but implanted within 5 au by the giant planet. Without accounting for the inhibiting effect of volatiles on melt production, over the timespan covered by our simulations this extended definition of target bodies raises the melt production by a few per cent in the case of Jupiter's in situ formation while in the case of Jupiter's migration it doubles the melt production. When the decrease in melt production is accounted for, however, the melt production in Jupiter's in situ scenario remains unchanged while it grows by 10\% in Jupiter's extended migration scenario (see also Main Text for discussion).

\subsection*{iSALE simulation}

Collisions between planetesimals are simulated by the iSALE shock-physics code\cite{iSALE}. The setup of the simulation is essentially the same as that in the previous study by \cite{Johnson}. The main difference is the porosity of the planetesimals. The parameters of the simulation are listed in Table~1. We adopted $0.4$ as the standard porosity value. Figure~ 1c shows the porosity dependence of the volume of the melt layer. The amount of melt increases as the porosity increases to $0.8$. High porosity leads to large energy dissipation owing to compaction, which results in the efficient production of melt.  We calculated the fraction of melt and silicate vapour from the entropy change\cite{Kurosawa}. The fraction of silicate vapour is negligible, and almost all the melt has a velocity exceeding the escape velocity of the target planetesimal. Other parameters were obtained from the previous study by \cite{Johnson}.

The melt production efficiencies from the iSALE simulations are computed considering head-on planetesimal collisions with a fixed size ratio of $4$. Melt production rates of oblique impacts would be lower than those of head-on collisions by a factor of 0.8\cite{Davison14}. Therefore, data in Fig.~1 may be overestimated by a factor of roughly $1.3$. Moreover, the ratio $M_{\rm melt}/M_{\rm imp}$ in $100-100$, $100-200$, $100-800$\,km diameter planetesimal collisions are 0.26, 0.45, and 0.56, respectively. This factor introduces another uncertainty of a factor of $2$. The most important factor arises from adding water in the collision, which can reduce the melt production by a factor of $10$\cite{Cashion21}.



\subsection*{Melt breakup simulation}

\subsubsection*{Basic equations}
Silicate melt with a spatial density $\rho_{\rm melt}(r)$ and velocity $v_{\rm melt}(r)$ is dragged by gas having a density $\rho_{\rm gas}(r)$ with $v_{\rm gas}(r)$, where $r$ is the distance from the center of the target planetesimal (Fig.~\ref{Sche}). The gas drag force is given by $\pi(D/2)^2C_{\rm D}\rho_{\rm gas}(v_{\rm melt}-v_{\rm gas})^2/2$, where $D$ is the droplet diameter and $C_{\rm D}$ is the gas drag coefficient\cite{gas}.  This gives the acceleration of the melt droplet as $m_{\rm melt}Dv_{\rm melt}/dt$, where $m_{\rm melt}=\pi D^3\rho_{\rm mat}/6$. ($\rho_{\rm mat}=2650\,{\rm kg\,m}^{-1}$ is the material density of the silicate melt\cite{Sakamaki}) Then the acceleration of the melt droplet is given by $Dv_{\rm melt}/dt=3C_{\rm D}\rho_{\rm gas}(v_{\rm melt}-v_{\rm gas})^2/(4\rho_{\rm mat}D)$. 
The equations of motion for the melt and gas components are given by
\begin{eqnarray}
  \label{meltmotion}
{\partial v_{\rm melt}\over \partial t}+v_{\rm melt}{\partial v_{\rm melt}\over \partial r}
&=& -\rho_{\rm gas}A(v_{\rm melt}-v_{\rm gas}) \\
{\partial v_{\rm gas}\over \partial t}+v_{\rm gas}{\partial v_{\rm gas}\over \partial r}
&=& -{1\over \rho_{\rm gas}}{\partial P\over \partial r}-\rho_{\rm melt}A(v_{\rm gas}-v_{\rm melt}),
\label{motion}
\end{eqnarray}
where $P$ is the gas pressure. The acceleration of gas comes from the backreaction, and the gas drag factor $A$ is given by 
\begin{equation}
  A={3C_{\rm D}|v_{\rm melt}-v_{\rm gas}|\over 4\rho_{\rm mat}D},
  \label{A}
\end{equation}
where $\rho_{\rm mat}=2650\,{\rm kg\,m}^{-3}$ is the material density of hydrous melt\cite{Sakamaki}. Because the Reynolds number is much larger than unity, $C_{\rm D}=0.44$ is adopted\cite{gas}.


The specific internal energy of the gas $e$,  and melt temperature $T_{\rm melt}$ change because of heat transfer between them as
\begin{eqnarray}
    \label{egas}
&& {\partial e\over \partial t}
  +v_{\rm gas}{\partial e\over \partial r}
  =-{P\over \rho_{\rm gas}}{1\over r^2}{\partial r^2v_{\rm gas}\over \partial r}
  -C_{\rm gas}{T_{\rm gas}-T_{\rm melt}\over \tau}
  +\rho_{\rm melt}A(v_{\rm melt}-v_{\rm gas})^2\\
&&   {\partial T_{\rm melt} \over \partial t}
  + v_{\rm melt}{\partial T_{\rm melt} \over \partial r} ={\rho_{\rm gas}C_{\rm gas}
    \over \rho_{\rm melt} C_{\rm melt}}{T_{\rm gas}-T_{\rm melt}\over \tau},
\end{eqnarray}
where $C_{\rm gas}=k/(m[\gamma-1])$ ($\gamma=4/3$ is the specific heat ratio of H$_2$O) and $C_{\rm melt}=2.3\times 10^3\,{\rm J\,kg^{-1}\,K^{-1}}$ are the specific heat of the gas and melt, respectively. Note that the specific heat of the melt includes the latent heat of fusion\cite{Neumann} ($4\times 10^5\,{\rm J\,kg^{-1}}$) divided by the temperature difference between the liquidus and solidus (400\,K). We assume that the transfer proceeds at a timescale $\tau$. Here, we adopt $\tau=1\,$s, which is the timescale of heat conduction within a  1\,mm silicate sphere. The last term on the right-hand side of Eq.~(\ref{egas}) is the heat produced by the friction between the gas and melt. 

The continuity equations for the melt and gas are respectively given by

\begin{eqnarray}
  {\partial \rho_{\rm melt}\over \partial t}+{1\over r^2}{\partial r^2\rho_{\rm melt}v_{\rm melt}\over \partial r}&=&0\\
{\partial \rho_{\rm gas}\over \partial t}+{1\over r^2}{\partial r^2\rho_{\rm gas}v_{\rm gas}\over \partial r}&=&0.
\end{eqnarray}

The size of a melt droplet is determined by breakup droplets due to gas flow. A droplet breaks up when the dynamic pressure of gas exceeds the pressure due to surface tension. The Weber number is defined as the ratio between these two quantities: ${\rm We}=\rho_{\rm gas} (v_{\rm gas}-v_{\rm melt})^2D/\sigma$, where $\sigma$ is the surface energy of the melt. If this number exceeds a critical value, the droplet breaks up into smaller droplets. The critical size for droplet breakup is given as:
\begin{equation}
  D_{\rm c}={{\rm We_c}\sigma \over \rho_{\rm gas} (v_{\rm gas}-v_{\rm melt})^2}.
  \label{Dmethod}
\end{equation}
The breakup occurred when the volume fraction of the melt was less than\cite{Sparks} $0.2$. After breakup, the size of the droplet was assumed to be $D=D_{\rm c}/2$\cite{Kadono}. The coalescence of the droplets could be disregarded. $D_{\rm c}$ evolves depending on the relative velocity between the gas and the melt. The size of the droplet was further reduced if $D_{\rm c}/2$ computed by Eq.~(\ref{Dmethod}) became smaller than the current droplet size.

Below the breakup limit, the gas exists as bubbles inside the melt. The melt layer expands as the bubbles grow. In this case, the expansion velocity of the melt and gas are almost the same as $v_{\rm melt}\simeq v_{\rm gas}$. Multiplying $\rho_{\rm melt}$ and $\rho_{\rm gas}$ to Eqs.~(\ref{meltmotion}) and (\ref{motion}), respectively, and adding them together leads to the equation describing the expansion of the mixture as
\begin{equation}
{\partial v_{\rm gas}\over \partial t}+v_{\rm gas}{\partial v_{\rm gas}\over \partial r}
= -{1\over \rho_{\rm melt}+\rho_{\rm gas}}{\partial P\over \partial r},
\end{equation}
 where $v_{\rm gas}\simeq v_{\rm melt}$ is the expansion velocity of the mixture. In the numerical simulation, we set $D=1\,\mu$m artificially to achieve the perfect coupling of the melt and gas.
      
We determined the droplet size as a function of $r$. The droplet size at particular $r$ changes even if breakup does not occur because of the advection due to gas flow. The advection of droplet size $D$ is expressed by
\begin{equation}
{\partial D\over \partial t}+v_{\rm melt}{\partial D\over \partial r}=0.
\end{equation}

The initial temperature of the melt layer $T_0$ is $1800\,$K, and volatile material with a mass fraction $f$ is uniformly distributed in the melt layer ($R_{\rm pla}\le r \le R_{\rm pla}+L_0$). The initial spatial gas and melt densities are $\rho_{\rm gas,0}=f\rho_{\rm mix}$ and $\rho_{\rm melt,0}=(1-f)\rho_{\rm mix}$, respectively,  where $\rho_{\rm mix}=\rho_{\rm mat}/(1-f+\rho_{\rm mat}f/\rho_{\rm H_2O})$ ($\rho_{\rm H_2O}=1000\,{\rm kg\,m^{-3}}$). The temperature and the volatile mass fraction far from the melt layer are set to $T_{\infty}=200$\,K and 0.99, respectively. The gas mass fraction of 0.99 corresponds to the standard gas/solid ratio of the interstellar medium (actually, the main component of the disk gas is H$_2$, not volatiles; here we neglect the difference for simplicity). The initial gas pressure $P_0$ is $\rho_{\rm gas,0}kT_0/m$. 

To avoid numerical numerical instabilities due to the discontinuity at the boundary between the planetesimal and the disk gas, the densities and gas pressure are assumed to decrease smoothly with a length scale of $L_0/10$. The distributions of the spatial densities of the gas $\rho_{\rm gas}(r)$ and melt $\rho_{\rm melt}(r)$, and the gas pressure $P(r)$  are respectively given by 
\begin{eqnarray}
  \rho_{\rm gas}(r) &=& (\rho_{\rm gas,0}-\rho_{\rm gas,\infty})e^{-10(r-R_{\rm pla}-L_0)/L_0}+\rho_{\rm gas,\infty}\\
  \rho_{\rm melt}(r)&=& (\rho_{\rm melt,0}-\rho_{\rm melt,\infty})e^{-10(r-R_{\rm pla}-L_0)/L_0}+\rho_{\rm melt,\infty}\\
  P(r) &=&(P_0-P_{\infty})e^{-10(r-R_{\rm pla}-L_0)/L_0}+P_{\infty}
  \end{eqnarray}
where $\rho_{\rm gas,\infty}$ and $\rho_{\rm melt,\infty}$ are the gas and melt densities far from the initial melt layer, respectively. Here, we adopt $\rho_{\rm gas,\infty}=2\times 10^{-7}\,{\rm kg\,m}^{-3}$ (=100$\rho_{\rm melt,\infty}$), $P_{\infty}=\rho_{\rm \infty}kT_{\infty}/m$. At $r=R_{\rm pla}$ (surface of the target planetesimal), the boundary conditions are given such that the gradient of all physical quantities is zero.

The above equations were numerically solved by Cubic-Interpolated Pseudo-Particle (CIP) method\cite{cip}. The radial distance $r$ is divided into incremental distance $dr(r)$, which logarithmically enlarges as $r$ increases. The location of the outer boundary is $r=3.6\times 10^5\,$km,  sufficiently far from the initial surface of the melt layer so that the expanding gas (and melt) do not reach the boundary during the simulation.

\subsubsection*{Semi-analytical solutions}
The three parameters in this simulation are the volatile mass fraction $f$, molecular weight $m$, and thickness  $L_0$ of the melt layer. The dependence of the size and the cooling rate on $f$, $m$, and $L_0$ can be derived analytically. The gas component cooled through expansion. The work done by the gas in a unit mass of the gas-melt mixture, which initially occupies a volume $V_0$ containing $f/m$ molecules, is written as $fkT/m\log (V/V_0)$, where $k$ is the Boltzmann constant and $V(t)$ is the volume at time $t$. At large times, gas expands spherically, and the ratio $V(t)/V_0$ can be given by $L(t)^3/(3R_{\rm pla}^2L_0)$, where $L(t)$ is the radius of the expanding gas. The expansion is driven by the pressure gradient $-\partial P/\partial x\rho_{\rm gas}$ in Eq.~(\ref{motion}). This acceleration can be approximated as $fc_0^2/L$, where $c_0=\sqrt{kT/m}$ is the sound speed of the gas, taking into account the gas mass fraction $f$. Then, the thickness of the melt layer at time $t$ is given by $L\simeq fc_0^2t^2/L$. From this equation, the thickness of the melt layer is approximated as $L(t)\simeq \sqrt{f} c_0t$ (the $\sqrt{f}$ dependence is confirmed by numerical experiments illustrated in the Supplementary Information). The temperature drop in the gas-melt mixture is then 
\begin{equation}
  \Delta T = f{kT_0\over mC_{\rm eff}}\log\left({(a\sqrt{f} c_0t)^3\over 3R_{\rm pla}^2L_0}\right),
  \label{DT}
\end{equation}
where $C_{\rm eff}=fC_{\rm gas}+(1-f)C_{\rm melt}$ ($C_{\rm gas}=k/(m[\gamma-1])$) is the specific heat of the gas-melt mixture, and the constant $a$ is introduced to correct the thickness of the melt layer from the above estimation to that determined from the numerical simulation. Differentiation of Eq.~(\ref{DT}) by $t$ gives a cooling rate of
\begin{equation}
  \left|{dT\over dt}\right| = {3fkT_0\over mC_{\rm melt}t}.
  \label{dTdt2}
  \end{equation}
The time $t_{\rm c}$ required for a temperature drop $\Delta T_{\rm c}$ is determined from Eq.~(\ref{DT}) as
\begin{equation}
  t_{\rm c} = {(3R_{\rm pla}^2L_0)^{1/3}\over a\sqrt{f} c_0}\exp \left({m\Delta T_{\rm c}C_{\rm eff}\over 3fkT_0}\right).
  \label{tDT}
  \end{equation}
From Eq.~(\ref{tDT}), the average cooling rate is written as
\begin{equation}
  \left|{dT\over dt}\right|_{\rm av} = {\Delta T_{\rm c}\over t_{\rm c}}= {a\sqrt{f}c_0\Delta T_{\rm c}\over (3R_{\rm pla}^2L_0)^{1/3}}\exp\left({-m\Delta T_{\rm c}C_{\rm eff}\over 3fkT_0}\right).
  \label{dTdt3}
  \end{equation}
By fitting the numerical result (Fig.~\ref{matome}d) with Eq.~(\ref{dTdt3}), the constant $a$ is determined as $a=3.34$. Using this $a$, Eq.~(\ref{dTdt3}) fits the numerical results well, as shown in Figs.~\ref{Fig3}b and \ref{matome}d, e, and f.

The acceleration of melt and gas (left-hand side of Eqs.~(\ref{meltmotion}) and (\ref{motion})) are almost the same. Because $\rho_{\rm melt} > \rho_{\rm gas}$, $\rho_{\rm gas}A(v_{\rm melt}-v_{\rm gas}) < \rho_{\rm melt}A(v_{\rm melt}-v_{\rm gas})$. Thus the two terms on the right-hand side of Eq.(\ref{motion}) can be approximated to be equal as
\begin{equation}
  -{1\over \rho_{\rm gas}}{\partial P\over \partial r}=\rho_{\rm melt}A(v_{\rm melt}-v_{\rm gas})=\rho_{\rm melt}{3C_{\rm D}{\rm We_c}\sigma\over 8\rho_{\rm mat}\rho_{\rm gas}D^2}.
  \label{Ap1}
\end{equation}
Because the breakup of the melt layer proceeds at the beginning of the expansion, the pressure gradient term can be approximated as $fc_0^2/L_0$ taking account of the gas mass fraction. Note that $\rho_{\rm gas}$ and $\rho_{\rm melt}$ are the spatial densities of the gas and droplets respectively, and $\rho_{\rm mat}$ is the material density of the droplets. Because both $\rho_{\rm gas}$ and $\rho_{\rm melt}$ decrease simultaneously, the ratio $\rho_{\rm melt}/\rho_{\rm gas}$ can be approximated by $\rho_{\rm mat}/\rho_{\rm H_2O}$. Equation~(\ref{Ap1}) is then rearranged as
\begin{equation}
  {fc_0^2\over L_0}
  ={3C_{\rm D}{\rm We_c}\sigma \over 8 \rho_{\rm H_2O}D^2}.
  \label{fc}
\end{equation}
Solving this equation with respect to $D$ gives
\begin{equation}
  D=
  b\left({3C_{\rm D}\sigma{\rm We_c}L_0m\over 8f\rho_{\rm H_2O}kT_0}\right)^{1/2},
  \label{D}
\end{equation}
where $b=0.32$ is an empirical factor to fit the numerical results. $b$ depends on the critical volume fraction of melt when the breakup occurs.

By inserting the values, the droplet diameter is obtained as 
\begin{equation}
  D=1.15
  \left({f\over 0.1}\right)^{-1/2}
   \left({m\over 18m_{\rm H}}\right)^{1/2}
  \left({L_0\over 10\,{\rm km}}\right)^{1/2}\,{\rm mm}.
  \label{Dfit2}
\end{equation}
Figures~\ref{matome}a, b, and c compare the numerical results and the analytical formula in Eq.~(\ref{Dfit2}). The numerical results are reproduced well by Eq.~(\ref{Dfit2}).




\newpage

\section*{Supplementary Information}

\subsection*{Dependence of the expansion velocity of the melt layer on the volatile mass fraction}

The expansion of the melt layer is driven by the expansion of the volatile material. The expansion velocity determines the cooling rate of the melt and the size of the droplets, and is therefore the key parameter in the semi-analytical solutions. The expansion velocity depends on the volatile mass fraction: when the volatile mass fraction increases so does the gas pressure gradient, while the mass of the melt droplets decreases. Fig.~\ref{FigS2}a shows the velocity distribution for different values of the volatile mass fraction $f$ at constant time $t=110$\,s, clearly illustrating both the dependence of the peak velocity from $f$ and how the peak velocity always exceeds the sound speed ($c_0\simeq 1\,{\rm km\,s^{-1}}$). Fig.~\ref{FigS2}b displays the expansion of the melt layer for the same values of the volatile mass fraction $f$. The slope of the curves represents the expansion velocity and immediately shows how it is almost constant until the cooling finishes (see Fig.5b in the Main Text).



Fig.~\ref{FigS2}c shows the dependence of the expansion velocity on the volatile mass fraction $f$. The dependence $dL/dt \propto f^{0.47}$ resulting from the numerical simulations is in agreement with the analytical estimation of $dL/dt\propto \sqrt{f}$. The expansion velocities are $1.4c_0$, $2.0c_0$, and $2.7c_0$ for $f=0.05$, $0.1$, and $0.2$, respectively. Note that the velocity is not produced by the impact itself but arises from the thermal expansion of the volatile material using the heat contained in the melt droplets.

The expansion velocity in panel c is smaller than the peaks shown in panel a. The peaks correspond to the shock wave surface. Because the shock wave travels much faster than the gas+melt mixture, the spatial density around the peak is far from the melt layer surface, where the spatial density of melt is that in the protoplanetary disk $\rho_{\rm melt,\infty}=2\times 10^{-9}\,{\rm kg\,m^{-3}}$, which is much smaller than the density of the melt layer $2.7\times 10^{-3}\,{\rm kg\,m^{-3}}$ at $t=1\,$hr (Fig.~5d: the initial density is $\rho_{\rm melt,0}=2.650\,{\rm kg\,m^{-3}}$). The melt layer surface $L(t)$ locates at $L(t)/R_{\rm pla}=1.5$, $1.8$, and $2.2$ for $f=0.05$, $0.1$, and $0.2$, respectively in Fig.~\ref{FigS2}a.


\subsection*{The inner Solar System after Jupiter's formation}

The dynamical and collisional environment responsible for the production of chondrules proves highly effective in altering the asteroid belt. At the beginning of our simulations the mass of solids contained in the orbital region of the asteroid belt between 2 and 3 au is about 2.4 M$_\oplus$. At the end of Jupiter's formation process and of the peak of chondrule production, a large fraction of this initial mass has been converted into collisional debris while planetesimals originating from beyond 3 au have been implanted into the orbital region of the asteroid belt. Specifically, in the Jovian in situ formation scenario the mass contained in the asteroid belt decreases to about 2 M$_\oplus$, of which about 0.9 M$_\oplus$ have been implanted from beyond 3 au. In the Jovian extensive migration scenario the final mass contained between 2 and 3 au drops to 0.6 M$_\oplus$, of which 0.3 M$_\oplus$ are implanted from beyond 3 au.

In the Jovian in situ formation scenario the implanted bodies originate from the now depopulated region between 3 and 5 au, while in the Jovian extensive migration scenario they originate from the region comprised between 3 and 30 au (see Fig. \ref{FigS1}). Both scenarios are globally compatible with the observed overlapping distributions of S and C type asteroids\cite{asteroid2015} but predict the implantation of different types and abundances of ices in the asteroid belt. The presence of ammoniated minerals on Ceres\cite{ceres} and other large main belt asteroids\cite{Kurokawa2022} appears to favour the Jovian extensive migration scenario due to the colder origin of its implanted bodies. The current constraints on the temperature profile of the solar nebula, however, do not permit to exclude Jupiter's in situ formation as they allow for colder disks where the NH$_3$ snowline falls between 3 and 4 au\cite{Oberg19}.

The collisional environment responsible for the production of chondrules has also implications for the ongoing planet formation process in the inner Solar System. Specifically, as the debris and pebbles are more affected by the disk gas than the primordial planetesimals, the resulting inward transport of material can promote the growth of surviving large planetesimals and existing planetary embryos during the lifetime of the solar nebula\cite{Turrini2023}, consistently with recent constraints on the accretion timescales of the terrestrial planets\cite{Lammer21}. The resulting gravitational interactions between the growing embryos and Jupiter are known to enhance and sustain the dynamical excitation of the inner solar system\cite{OBrien07,Johnson}, hence the chondrule production process, beyond the 0.5 Myr caused by Jupiter alone while triggering the depletion of the asteroid belt\cite{OBrien07,Johnson}. 

{Additional processes that can sustain and extend the duration of the chondrule production, alone or in conjunction with the role of planetary embryos, are the formation of Saturn  (\cite{Coradini2011,Turrini12,Ronnet2018}, see Main Text) and the dynamical excitation caused by the sweeping secular resonances during the dissipation of the solar nebula after the formation of the giant planets\cite{nagasawa2000}. Alternatively, the depletion of the asteroid belt and the extension of the chondrule production can be achieved by further inward migration of Jupiter as that invoked to explain the small mass of Mars\cite{tacky}.

Finally, these results have implications also for the interpretation of the origins of CB chondrules. CB chondrules were formed $\sim3.8$\,Myr after CAIs\cite{Wolfer2023} and considered as the products of planetesimal collisions\cite{Fedkin}. A detailed analysis of CB chondrules revealed that collisions with high water content ($\sim 20$\,\%) and a high cooling rate match their zoning profiles. This is consistent with our results on the $f$ dependence of the cooling rate (Fig.~4d in the Main Text) 
and the dynamical implantation of volatile-rich planetesimals in the inner Solar System. Their age, however, is inconsistent with them being associated with Jupiter's formation as previously proposed\cite{Johnson2016}, and indicates instead that they are the product of the formation of Saturn (see Main Text) or of the later collisional environment responsible for the depletion of the asteroid belt.

\newpage


  


\section*{Acknowledgements}
We gratefully acknowledge the developers of iSALE, especially G. Collins, K. W\"{u}nnemann, D. Elbeshausen, J. Melosh, and B. Ivanov. We deeply thank for constructive comments by anonymous reviewers. S.S. appreciates fruitful discussion with Dr. Nagasawa. D.T. wishes to thank Guy Consolmagno, Josep Trigo-Rodríguez, Leonardo Testi, Eugenio Schisano, Elenia Pacetti, Sergio Molinari, Danae Polychroni and Simona Pirani for the insightful discussions on meteoritics, protoplanetary disks and giant planet formation. This work was supported {by JSPS KAKENHI Grant Number 25K07383}, by the Italian Space Agency through ASI-INAF contract 2016-23-H.0 and 2021-5-HH.0 and by the European Research Council via the Horizon 2020 Framework Programme ERC Synergy “ECOGAL” Project GA-855130. Melt breakup and iSALE simulations were conducted on a PC cluster at the Center for Computational Astrophysics, National Astronomical Observatory of Japan. The n-body simulations with Mercury-Ar$\chi$es and the collisional analysis of the excited planetesimal disk with Debris were performed using the \emph{Genesis} cluster at INAF-IAPS and the computational support of Romolo Politi, John Scige Liu and Sergio Fonte is gratefully acknowledged.

\section*{Author contributions statement}
S. S. conceived the idea that Jovian formation could induce high-velocity collisions of planetesimals that can result in chondrule formation. D. T. performed the planetesimal excitation simulations with the Mercury-Ar$\chi$es code and evaluated the collisional production of chondrules with the Debris code. S. S. performed the iSALE  and melt expansion simulations. Both authors contributed to the preparation of the manuscript and the conclusions presented in this work.

\section*{Data Availability}
Data and programs are provided by a request to the corresponding author (S.S.).


\newpage 

\begin{figure}[t]
  \includegraphics[width=\linewidth]{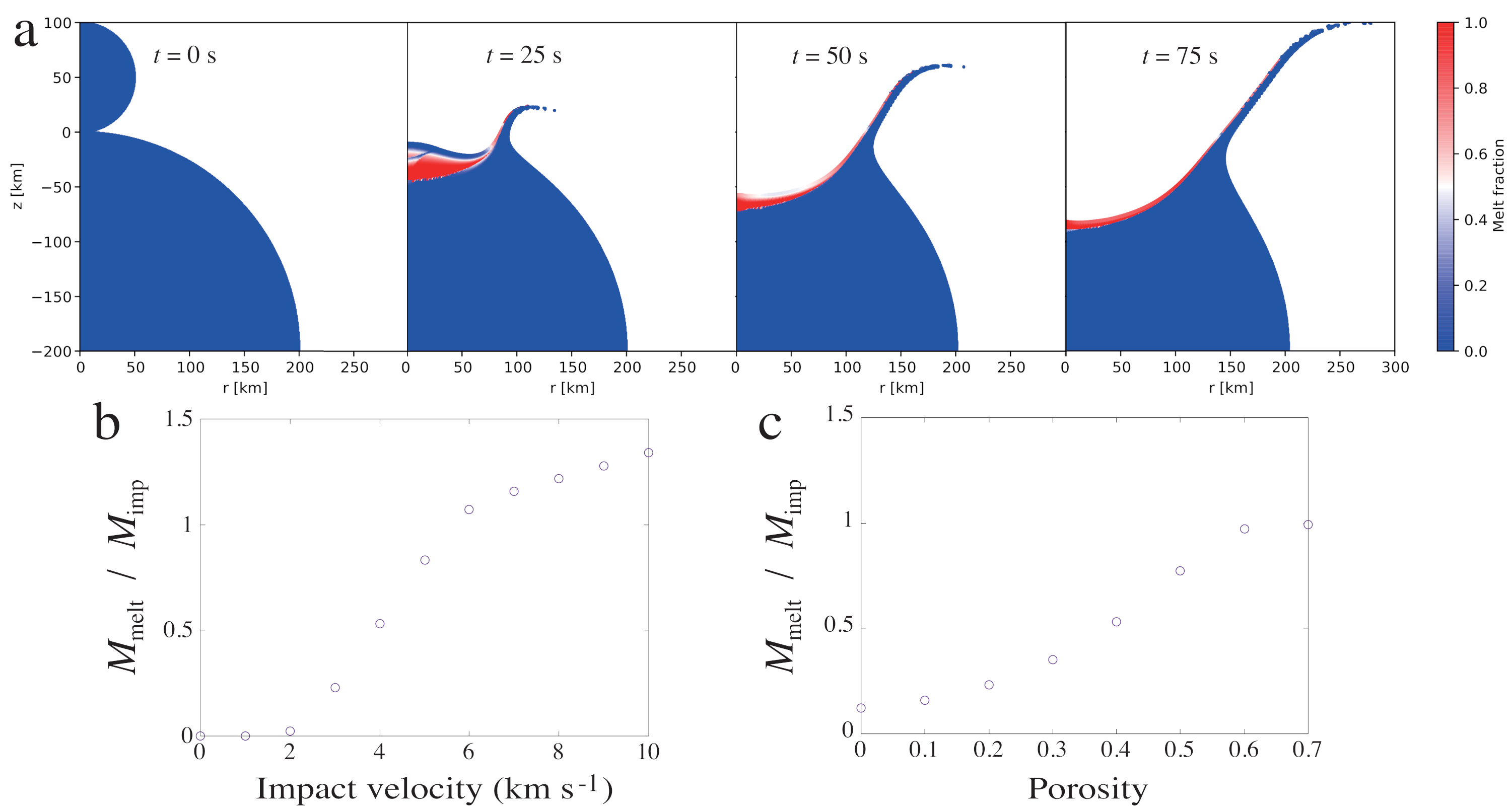}
 \caption{a: Snapshots of a head-on collision  (impact velocity, 5\,km\,s$^{-1}$) between $100$ and $400$\,km diameter planetesimals composed of dunite. Initial porosities of the 100 and 400\,km planetesimals are 0.4 and 0, respectively. From left to right, collision after $0$, $25$, $50$, and $75$\,s are shown. Cylindrically symmetrical coordinates where the vertical ($z$) axis is the center of symmetry  surrounded by the horizontal ($r$) axis are adopted. The color shows the volume fraction of the melt. The amount of melt reaches a maximum when $t=75$\,s, at which the melt thickness is $10$\,km at the central axis. b: The amount of the mass of melt normalized by the impactor mass $M_{\rm melt}/M_{\rm imp}$ as a function of the impact velocity with a fixed porosity of $0.4$. If the impact velocity is faster than $6\,{\rm km\,s}^{-1}$, the melt mass exceeds the mass of the impactor. c: The normalized amount of melt $M_{\rm melt}/M_{\rm imp}$ as a function of impactor porosity  with a fixed impact velocity of $4\,{\rm km\,s^{-1}}$}. 
\label{snap}
\end{figure}

\vspace{10cm}

\newpage

\begin{figure}[t]
\centering
\includegraphics[width=\textwidth]{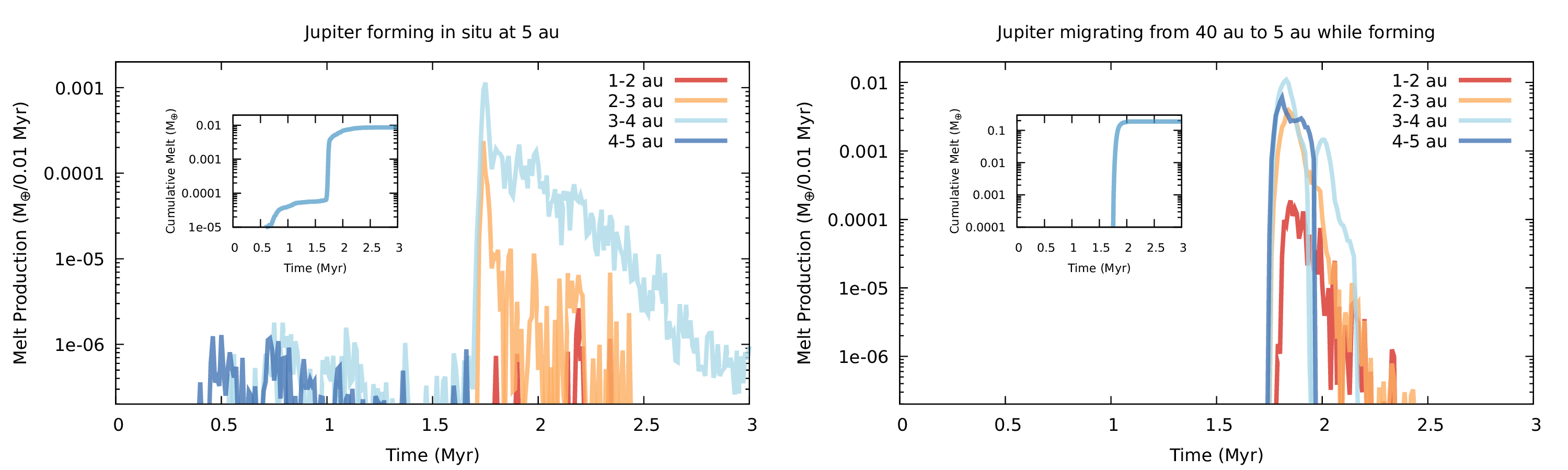}
\caption{Temporal evolution of the collisional production rate of chondrules across the different orbital regions of the inner Solar System. The collisional melt production in each annular region is computed by integrating overall impacts that occurred within its boundary in each temporal interval over which we resolve the collisional evolution (see Methods). The left plot shows the scenario of the in situ formation of Jupiter, and the right plot shows the scenario of extensive migration of Jupiter following its formation beyond the N$_2$ snowline. The peak efficiency in melt production is achieved between 3 and 4 au in both scenarios. The insets in the plots show the cumulative production of chondrules over time, the sharp increase at about 1.8 Myr is the result of Jupiter's runaway gas growth.}
\label{Fig1}
\end{figure}

\newpage
 
\begin{figure}[t]
  \includegraphics[width=\textwidth]{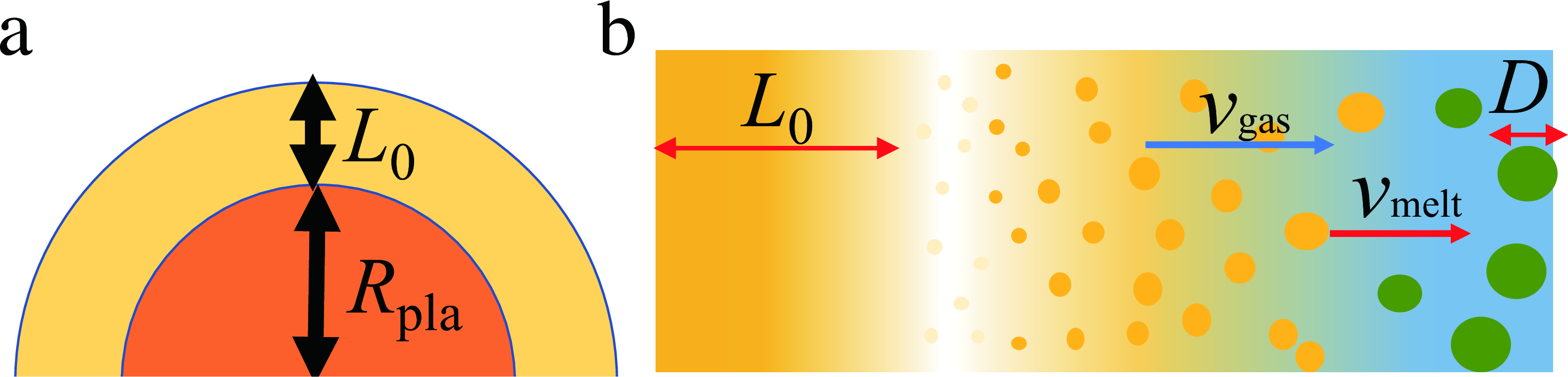}
  \caption{a: Schematic of the geometry of numerical simulation. The initial arrangement of the melt layer is approximated by the layer of thickness $L_0$ covering a planetesimal of radius $R_{\rm pla}$. b: Schematic of the melt layer expansion. Volatile material in the layer with thickness $L_0$ evaporates and expands with a velocity $v_{\rm gas}(r)$. Silicate melt is dragged by gas and has a velocity $v_{\rm melt}(r)$. The melt forms droplets of diameter $D$ determined by Weber number at which collisional equilibrium occurs. 
}
 \label{Sche}
 \end{figure}

\newpage

\begin{figure}[h]
   \vspace{-0cm}
  \centering
  \includegraphics[width=\textwidth]{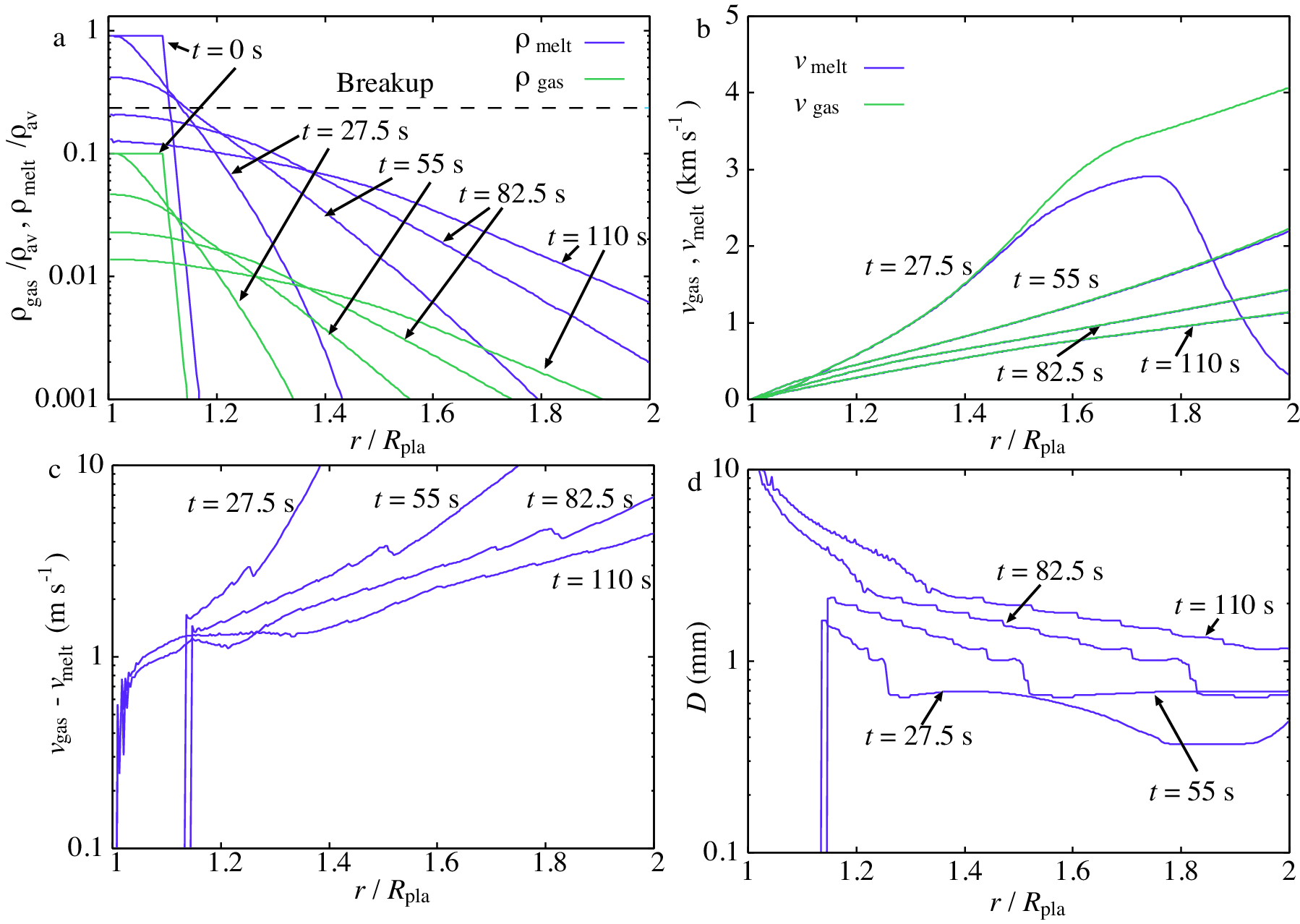}
  \vskip 0cm
  \caption{Results of melt breakup simulation using parameters of $f=0.1$, $L_0=10$\,km and $m/m_{\rm H}=18$. The horizontal axis is the distance $r$ from the center of the target planetesimal normalized by the radius of the target planetesimal $R_{\rm pla}$.  Snapshots of $t/t_0=27.5$, $55$, $82.5$, and $110$\,s are shown. a: evolution of spatial densities of melt ($\rho_{\rm melt}$: purple) and gas ($\rho_{\rm gas}$: green) components normalized by the initial average density of melt-gas mixture $\rho_{\rm av}$. The dashed horizontal line is the breakup density, where the volume fraction of melt is $0.2$. The initial surface of the melt layer is $r/R_{\rm pla}=1.1$. b: evolution of melt ($v_{\rm melt}$) and gas ($v_{\rm gas}$) velocities. The melt component is dragged by gas and $v_{\rm gas}> v_{\rm melt}$. c: evolution of the velocity difference between gas and melt $v_{\rm gas}-v_{\rm melt}$. d: Evolution of the distribution of the melt diameter $D$. The curve starts at $r/R_{\rm pla}=1.1$ when $t=27.5$ and $55\,$s because the breakup does not proceed at these periods. 
}
 \label{Fig3}
\end{figure}

\newpage

\begin{figure}[t]
  \includegraphics[width=\textwidth]{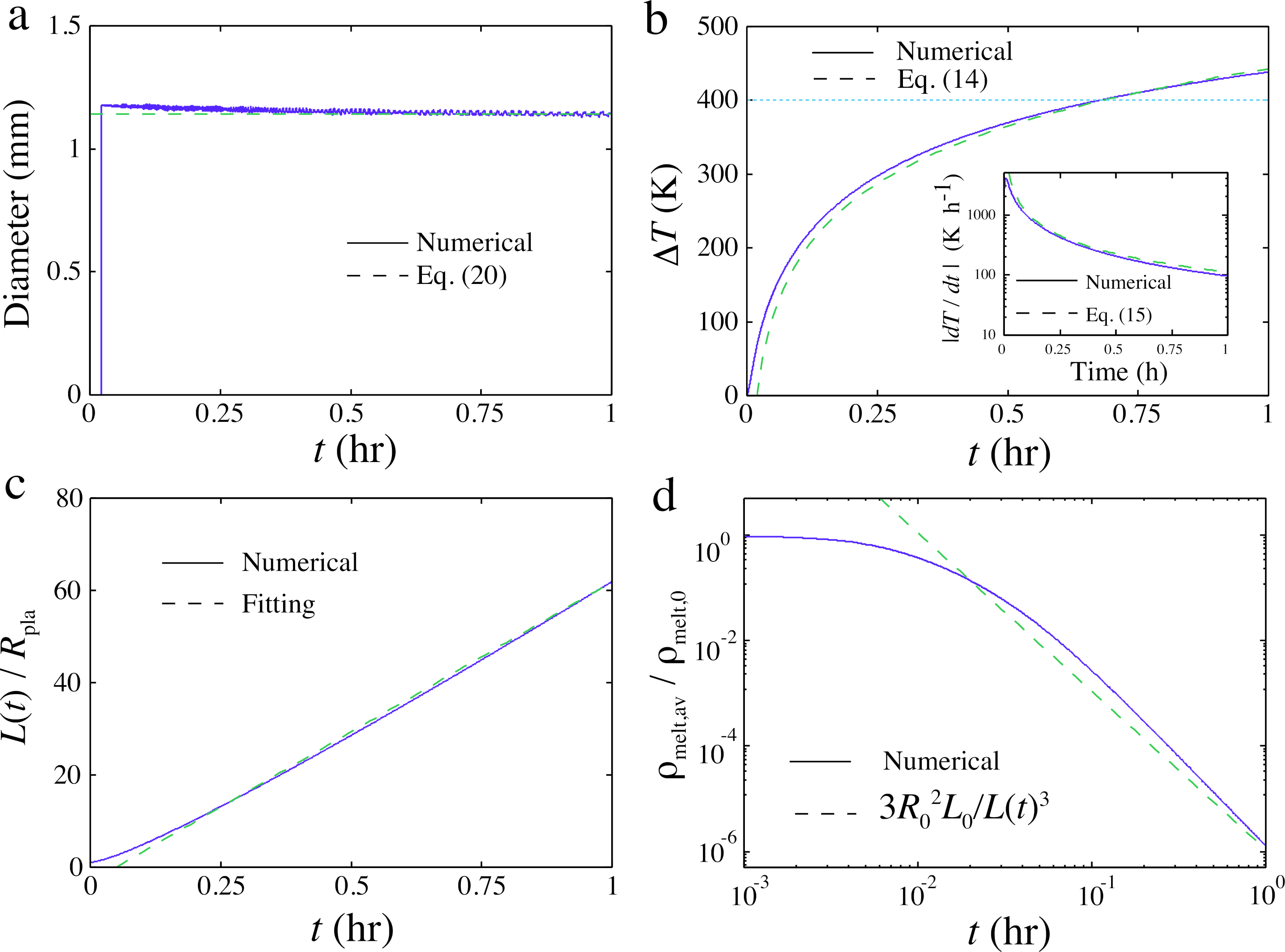}
  \caption{a: Evolution of average droplet size for simulation using $f=0.1$, $L_0=10\,$km, and $m/m_{\rm H}=18$ case. The numerical result (solid) and analytical solution (dashed) (Eq.(\ref{Dfit2}) are shown. The breakup of melt occurs ($t/t_0=75$) when the melt volume fraction falls below\cite{Sparks} $0.2$. b: Evolution of temperature decrease. The solid (dashed) lines are the numerical (analytical: Eq.~(\ref{DT})) results. The horizontal dotted line is $400$\,K, where the melt solidifies. Inset: evolution of the cooling rate. Solid (dashed) lines are the numerical (analytical: Eq.~(\ref{dTdt3})) results. c: Evolution of the location of the melt layer surface normalized by the planetesimal radius $L(t)/R_{\rm pla}$. The dashed line is the least-square fitting. d: Evolution of the average melt spatial density within $L(t)$ normalized by the initial melt density $\rho_{\rm melt,av}/\rho_{\rm melt,0}$ ($\rho_{\rm melt,0}=2047\,{\rm kg\,m}^{-3}$). The dashed line is an approximated formula $3R_0^2L_0^2/L(t)^3$, where $L(t)$ is given by the least-square fitting line in panel c.   
}
 \label{sizeevo}
 \end{figure}

\vspace{10cm}

\newpage

\begin{figure}[t]
  \includegraphics[width=\textwidth]{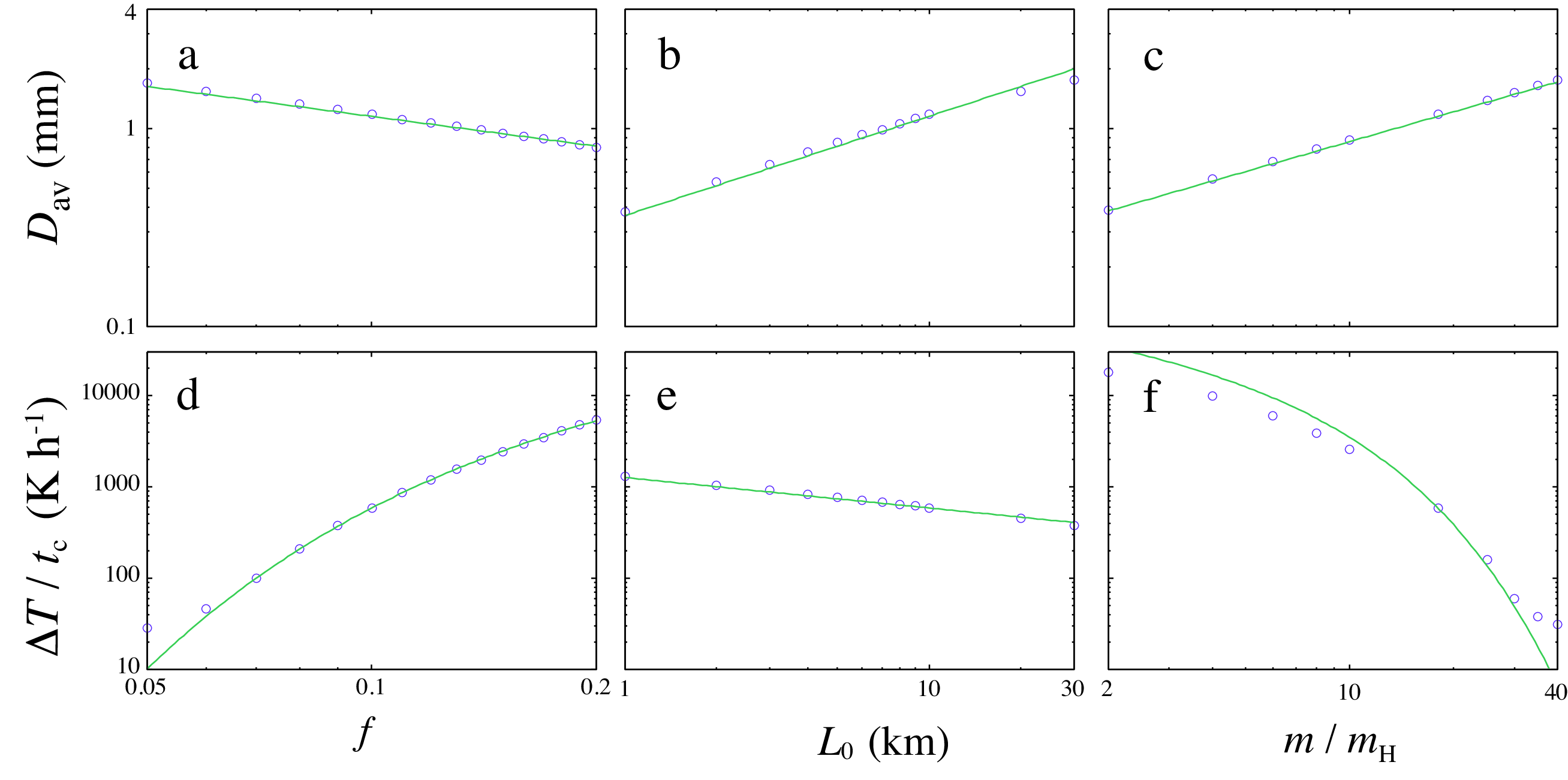}
 \caption{Parameter dependencies of droplet size $D$ (a--c) and cooling rate $dT/dt$ (d--f). a, d: Volatile mass fraction $f$ dependence. b, e: Melt layer  thickness $L_0$ dependence.  c, f: Molecular weight $m$ dependence. Solid lines are the semi-analytical equations given by Eqs.~(\ref{dTdt3}) and (\ref{Dfit2}). Each parameter is changed from its standard values ($f=0.1$, $L_0=10\,$km, and $m=18\,{\rm m_H}$) while keeping the other two kept constant). The initial temperature is fixed at $T_0=1800\,$K.}
 \label{matome}
 \end{figure}

\begin{table}
  \begin{tabular}{cc}
    Equation of state & ANEOS dunite \\
     Solidus temperature & $1373$\,K \\
    Porosity model & WUNNEMA \\
    Poisson's ratio & $0.25$ \\
    Simon A parameter & $1.52$\,GPa \\
    Simon C parameter & $4.05$ \\
Ivanov A parameter & $1.0\times 10^{-4}$ \\
Ivanov B parameter & $1.0\times 10^{-11}$ \\
Ivanov C parameter & $3.0\times 10^{8}$ \\
Sound speed ratio & $1.0$  \\
Initial temperature & $200$\,K \\
Size of high resolution cell  & $1$\,km \\
Number of high resolution cells horizontal direction & $600$ \\
Number of high resolution cells vertical direction & $1200$ \\
Strength at inifinite pressure YLIMINT& $3.26$\,GPa \\
Strain at which porous compaction begins EPSE0 & $0.01$\\
Strength of damaged material YDAM0& $10$\,kPa \\
Cohesion of damaged material YINT0& $5.07$\,MPa \\
Friction coefficient of damaged material FRICDAM& $0.63$ \\
Rate of porous compaction KAPPA& $0.98$ \\
Initial damage & $1.0$ \\
\end{tabular}
\caption{Parameters used in iSALE simulations.}
\end{table}

\newpage

\setcounter{figure}{0}
\renewcommand{\figurename}{Fig.}
\renewcommand{\thefigure}{S\arabic{figure}}

\begin{figure}[h]
   \includegraphics[width=\textwidth]{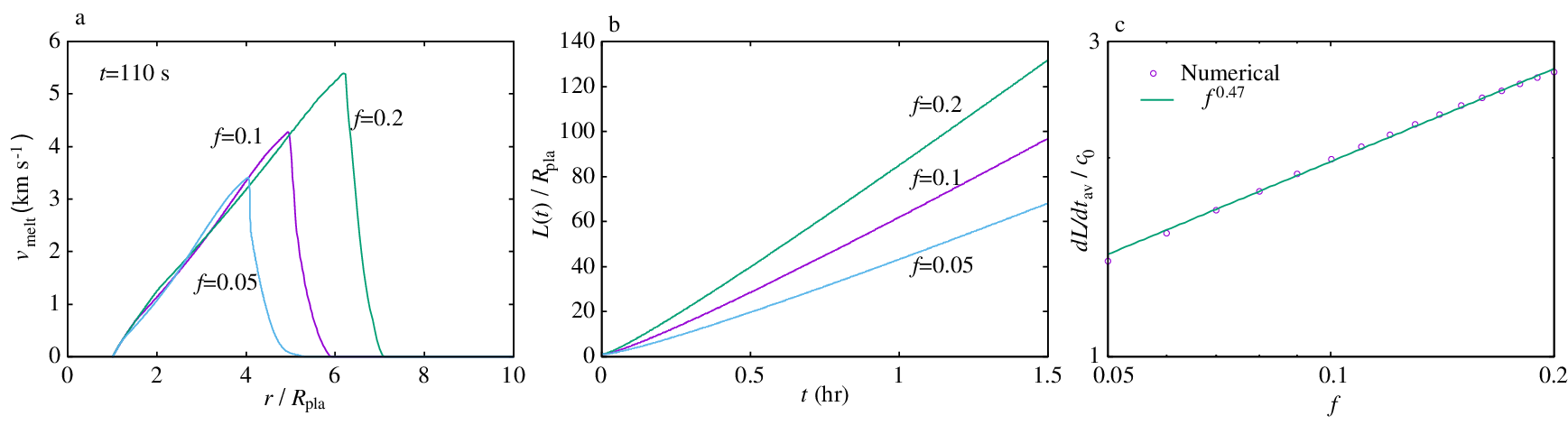}
   \caption{Expansion of the melt layer. The planetesimal radius $R_{\rm pla}$ is $100\,$km and the initial melt layer thickness $L_0$ is $10\,$km. a: Velocity distribution of melt $v_{\rm melt}(r,t)$ for different volatile mass fraction $f$. b: Evolution of the melt layer thickness $L(t)/R_{\rm pla}$ normalized by the planetesimal radius $R_{\rm pla}$ for different volatile mass fraction $f$. The average expansion velocity $dL/dt_{\rm av}$ is obtained by the least-square fitting of the curve. c: the normalized average expansion velocity of the melt layer $(dL/dt_{\rm av})/c_0$ normalized by the sound velocity $c_0=\sqrt{kT_0/m_{\rm H_2O}}\simeq 1\,{\rm km\,s^{-1}}$ as a function of the volatile mass fraction $f$. The average velocity was computed in the range of $0\le t\le 1.5$\,hr. The solid line is $dL/dt_{\rm av}\propto f^{0.47}$.}
 \label{FigS2}
\end{figure}

\begin{figure}[t]
 \includegraphics[width=\textwidth]{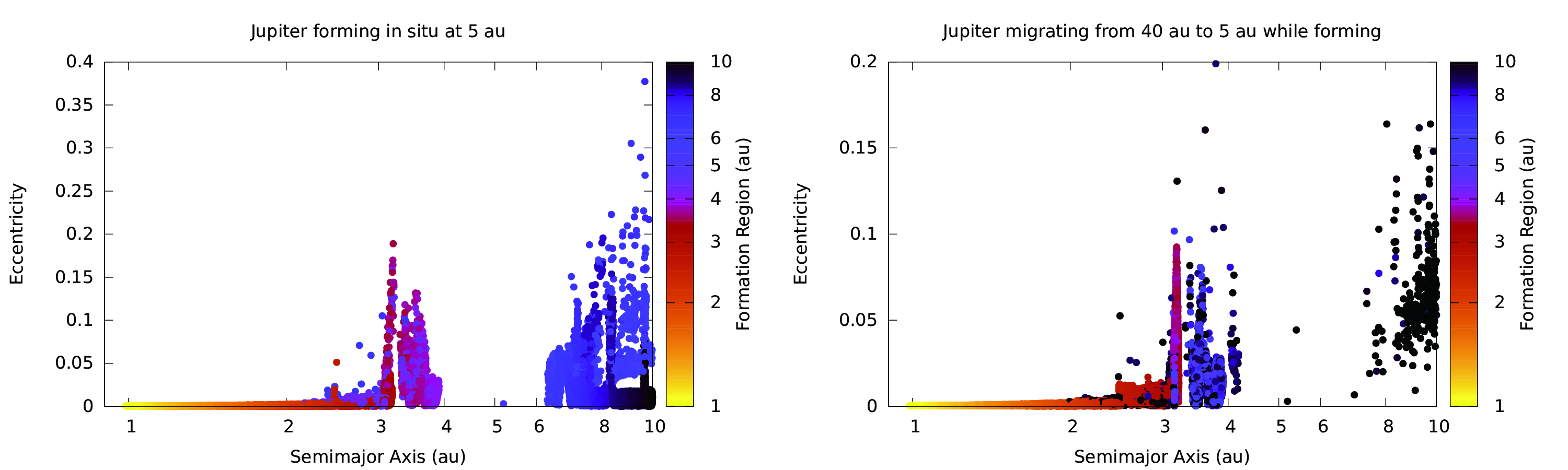}
 \caption{The solar nebula after Jupiter's formation. The left plot shows the dynamical state of the planetesimal disk in the solar nebula in the in situ formation scenario, the right plot in the extensive migration scenario. The semimajor axes are plotted in logarithmic scales to zoom on the inner Solar System. The formation region of the planetesimals is indicated by the color scale, with the black color identifying planetesimals formed at or beyond 10 au. Planetesimals from beyond the water snowline are systematically implanted in the orbital region of the current asteroid belt between 2 and 3 au.}\label{FigS1}
\end{figure}

\end{document}